\documentstyle[preprint,aps,epsfig]{revtex}
\tighten
\begin{document}
\draft
%__________________________________________________
%\twocolumn
%_______________________ Title, Authors ____________________________________
\preprint{\parbox[t]{85mm}{Preprint Numbers: \parbox[t]{48mm}
 {ANL-PHY-8214-TH-95 \\ UNITU-THEP-15/1995\\  hep-ph/9510284}}}
\title{Calculation of the anomalous $\gamma \pi ^* \to \pi \pi $ form factor}
\author{Reinhard Alkofer\footnotemark[1] and Craig D. Roberts\footnotemark[2]}
\address{\footnotemark[1]Institut f\"{u}r Theoretische Physik,
Universit\"{a}t T\"{u}bingen, \\
Auf der Morgenstelle 14, D-72076 T\"{u}bingen, FRG\\
\footnotemark[2]Physics Division, Argonne National Laboratory,\\
Argonne, Illinois 60439-4843, USA}
\maketitle
%-------------------------------------------------------------------
\begin{abstract}
The form factor for the anomalous process $\gamma \pi ^* \to \pi \pi$,
$F^{3\pi}(s,t,u)$, is calculated as a phenomenological application of the QCD
Dyson-Schwinger equations. The chiral-limit value dictated by the
electromagnetic, anomalous chiral Ward identity, $F^{3\pi}(0,0,0)= e
N_c/(12\pi^2 f_\pi^3)$, is reproduced, {\it independent} of the details of
the modelling of the gluon and quark 2-point Schwinger functions.  Using a
parametrisation of the dressed $u$-$d$ quark 2-point Schwinger function that
provides a good description of pion observables, such as $\pi$-$\pi$
scattering-lengths and the electromagnetic pion form factor,
$F^{3\pi}(s,t,u)$ is calculated on a kinematic range that proposed
experiments plan to explore.  Our result confirms the general trend of other
calculations; i.e., a monotonic increase with $s$ at fixed $t$ and $u$, but
is uniformly larger and exhibits a more rapid rise with $s$.
\end{abstract}

{\bf 1. Introduction.}  Hadronic processes involving an odd number of
pseudoscalar mesons are of particular interest because they are intimately
connected to the anomaly structure of QCD. The decay $\pi^0\to\gamma\gamma$
is the primary example of such an anomalous process.  That such processes
occur in the chiral limit ($m_\pi^2=0$) is a fundamental consequence of the
quantisation of QCD; i.e., of the non-invariance of the QCD measure under
chiral transformations even in the absence of current quark
masses\cite{Fuji}.  The $\pi^0\to\gamma\gamma$ decay rate can be calculated
from a quark triangle diagram and agreement with the observed rate requires
that the number of colours, $N_c$, equal three.  The transition form factor
for the related process $\gamma^\ast \pi^0 \to\gamma$ can be measured
experimentally\cite{AGN94} and has attracted keen theoretical
interest\cite{FMRP95,ABBC92} because it involves only one hadronic bound
state and provides a good test of QCD-based models and their interpolation
between the soft and hard domains.

Another anomalous form factor, accessible to experiment, is that which
describes the transition $\gamma \pi ^* \to \pi \pi $, denoted by
$F^{3\pi}(s,t,u)$.  This provides additional constraints on QCD based-models
because three hadronic bound states are involved.  The form factor
$F^{3\pi}(s,t,u)$ has been measured at Serpukhov in the Primakov reaction
$\pi^- A \to \pi^{-\prime} \pi^0 A^\prime$.\cite{Antipov} In this experiment
the considerable uncertainty in both the kinematic range and result make it
difficult to draw a conclusion regarding the accuracy of the theoretical
prediction for the chiral limit value of $F^{3\pi}(0,0,0)$\cite{ALTZ71}.  New
experiments are planned at CEBAF:\cite{Exp1} $\gamma p \to \pi^+ \pi^0 n$,
$s/m_\pi^2\in [4,15]$; and at FermiLab\cite{Exp2} via the Primakov reaction
using a 600 GeV pion beam, $s/m_\pi^2 \in [4,6]$.

\begin{figure}
  \centering{\
     \epsfig{figure=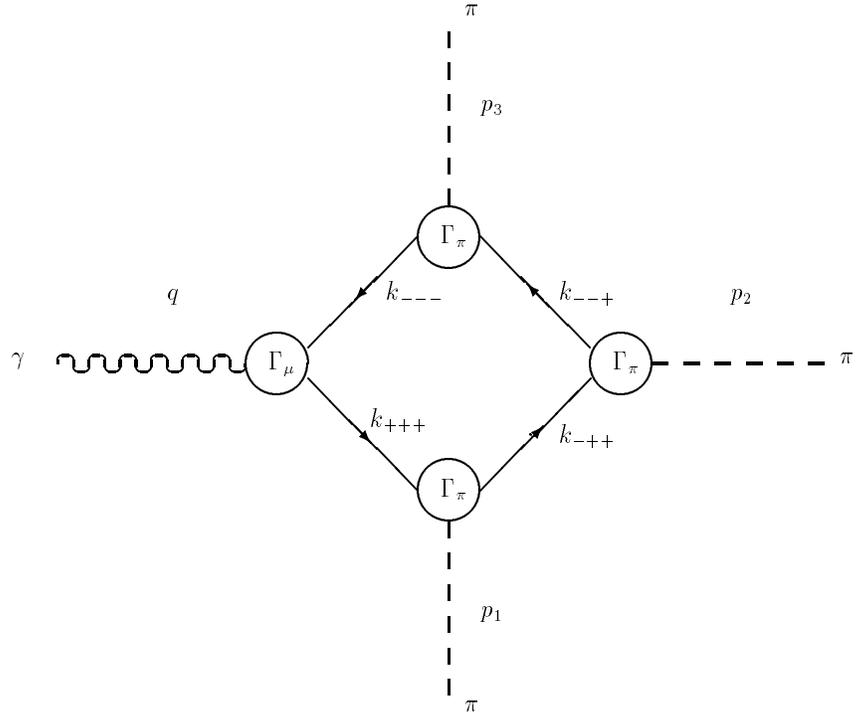,height=25cm,rheight=14cm}  }
\caption{This figure is a pictorial representation of the amplitude
identified with the $\gamma \pi ^* \pi \pi $ vertex.  The straight broken
external lines represent the outgoing $\pi$, the circles at the $\pi$ legs
represent the $\langle\pi|\overline{q}q\rangle$ Bethe-Salpeter amplitudes,
$\Gamma_\pi$, the wiggly line represents the photon, $\gamma$, the circle at
the $\gamma$ leg represents the regular part of the dressed quark-photon
vertex, $\Gamma_\mu$, and the full internal lines represent the dressed quark
2-point Schwinger function, $S$.
\label{figGPPP} }
\end{figure}

Herein we report a calculation of $F^{3\pi}(s,t,u)$ on the kinematic range to
be explored in the CEBAF experiment\cite{Exp1}.  We evaluate the amplitude
indicated by the diagram in Fig.~\ref{figGPPP}, in which the quark 2-point
Schwinger function, quark-photon vertex and quark-pion vertex (pion
Bethe-Salpeter amplitude) are dressed quantities whose form follows from the
extensive body of nonperturbative, semi-phenomenological Dyson-Schwinger
equations studies in QCD.\cite{DSErev,R94} In this way our calculation
provides for an extrapolation of the known large spacelike-$q^2$ behaviour of
these QCD Schwinger functions to the small spacelike-$q^2$ region, where they
are unknown and confinement effects are manifest.  This facilitates an
exploration of the relationship of physical observables to the
nonperturbative, infrared behaviour of these Schwinger functions.

This calculation employs the model forms introduced in Ref.~\cite{R94}.  The
quark 2-point Schwinger function has no Lehmann representation and hence may
be interpreted as describing a confined particle since this feature is
sufficient to ensure the absence of quark production thresholds in $S$-matrix
elements describing colour-singlet to singlet transitions.  The quark-photon
vertex, which describes the coupling of a photon to a dressed quark, follows
from extensive QED studies\cite{DMR94,BP94} and satisfies the Ward-Takahashi
identity.  This necessarily entails that the amplitude is current conserving.
In the chiral limit the quark-pion vertex is completely determined by the
quark 2-point Schwinger function\cite{DS79}, which is the manifestation of
Goldstone's theorem in this approach.  The extension to finite quark mass
requires a minimal modification and preserves Dashen's
relation\cite{Dashen69}.

The quark 2-point Schwinger function is parametrised such that it provides an
good description of $\pi$-$\pi$ scattering at, and to approximately $600$~MeV
above, threshold and the electromagnetic pion form factor at spacelike
momentum transfer.  No adjustment of the parameters is made in calculating
$F^{3\pi}(s,t,u)$.  The $\gamma \pi^\ast \to \pi\pi$ process probes the pion
bound state amplitude well outside the domain of the complex plane on which
it has been fitted; i.e., well into the timelike region, and hence provides a
new test of the model.  Therefore, interpreted within our framework, the
experimental determination of $F^{3\pi}(s,t,u)$ provides important new
information about the structure of the nonperturbative pion bound state
amplitude.

{\bf 2. The amplitude for \mbox{\boldmath $\gamma \pi ^* \to \pi \pi $}.}  In
Euclidean space, with metric \mbox{$\delta_{\mu\nu}={\rm diag}(1,1,1,1)$} and
$\gamma_\mu$ = $\gamma_\mu^\dagger$, the amplitude for the $\gamma \pi ^* \pi
\pi $ vertex (see Fig.\ \ref{figGPPP}) is
\begin{eqnarray}
\lefteqn{i\epsilon_{\mu\nu\rho\sigma}p_{1\nu}p_{2\rho}p_{3\sigma}
F^{3\pi}(p_1,p_2,p_3) =  } \nonumber \\
& & 2eN_c \int\case{d^4k}{(2\pi)^4}\,
   {\rm tr}_D \left[\Gamma_\mu(k_{---},k_{+++}) S(k_{+++})
         \gamma _5 \Gamma_\pi (k_{0++}) S(k_{-++})\right. \nonumber \\
 &&  \left.    \gamma _5 \Gamma_\pi (k_{-0+}) S(k_{--+})
         \gamma _5 \Gamma_\pi (k_{--0}) S(k_{---}) \right] ~.
\label{F}
\end{eqnarray}
In this expression the colour and flavour traces have been evaluated, leaving
only the Dirac trace, and we have employed the definition
\mbox{$
k_{\alpha\beta\gamma} = k + \case{\alpha}{2}p_1 + \case{\beta}{2}p_2
                          + \case{\gamma}{2}p_3$}.
Note that due to momentum conservation the photon momentum is equal to the
sum of pion momenta: $q=p_1+p_2+p_3$. In the following we adopt the
convention that the pions labelled 1 and 2 are on--shell whereas pion 3 is
off--shell.  The dressed quark--photon vertex is denoted by
$\Gamma_\mu(k_1,k_2)$, the pion Bethe--Salpeter amplitude by $\gamma _5
\Gamma_\pi (k)$ and the dressed quark 2-point Schwinger function by $S(k)$.
Equation~(\ref{F}) can be derived as an application of the formalism
described in Ref.~\cite{FT94}.  Evaluated with a dressed quark-photon vertex
that satisfies the Ward-Takahashi identity, which is a minimal requirement,
this expression is current conserving.

The quark 2-point Schwinger function can be written
\begin{eqnarray}
\label{Sp}
S(p)& =& - i \gamma\cdot p\, \sigma_V(p^2) + \sigma_S(p^2)\\
\label{Sinv}
&=& \frac{1}{i \gamma\cdot p\, A(p^2)  + m + B(p^2)}~,
\end{eqnarray}
where $m=m_u=m_d$ is the current quark mass, and can be obtained by solving
the quark Dyson--Schwinger equation (DSE)\cite{DSErev}.  The many studies of
this equation ensure that the qualitative features of the functions
$\sigma_S$ and $\sigma_V$ are well known for real, spacelike-$p^2$.

In Ref.~\cite{R94}, in order to avoid the need for a numerical solution of
the quark DSE, the following {\it approximating} algebraic forms are used:
\begin{eqnarray}
\label{SSM}
\bar\sigma_S(x) & = & C_m e^{-2 x} +
        \frac{1 - e^{- b_1 x}}{b_1 x}\,\frac{1 - e^{- b_3 x}}{b_3 x}\,
        \left( b_0 + b_2 \frac{1 - e^{- \Lambda x}}{\Lambda\,x}\right)
        + \frac{\bar m}{x + \bar m^2}
                \left( 1 - e^{- 2\,(x + \bar m^2)} \right)\\
\label{SVM}
\bar\sigma_V(x) & = & \frac{2 (x+\bar m^2) -1
                + e^{-2 (x+\bar m^2)}}{2 (x+\bar m^2)^2}
                - \bar m\, C_m\, e^{-2 x}.
\end{eqnarray}
with $k^2= 2\,D\,x$, $\bar\sigma_S(x) = \sqrt{2\,D}\,\sigma_S(k^2)$, $\bar
\sigma_V(x) = 2\,D\,\sigma_V(k^2)$ and $\bar m = m/\sqrt{2\,D}$.  The quantity
$\lambda=\sqrt{2\,D}$ is a mass--scale related to the infrared behavior of
the gluon 2-point Schwinger function\cite{DSErev}.

At large spacelike-$p^2$ the model forms behave as
\begin{eqnarray}
\label{SUV}
\sigma_S(p^2) \approx \frac{m}{p^2} - \frac{m^3}{p^4}
+ \frac{b_0}{b_1 b_3} \frac{\lambda^3}{p^4} + \ldots \;\;
& \; \mbox{and} \; & \;\;
\sigma_V(p^2) \approx \frac{1}{p^2} - \frac{D+m^2}{p^4} + \ldots ~,
\end{eqnarray}
whereas in QCD one has
\begin{equation}
\sigma_S(p^2) \approx \frac{\hat{m}}{p^2
\left[\case{1}{2}\ln\left(p^2/\Lambda_{\rm QCD}^2\right)\right]^d}
- \frac{4\pi^2 d}{3} \frac{\langle\bar q q\rangle}
{p^4\left[\ln\left(p^2/\Lambda_{\rm QCD}^2\right)\right]^{d-1}} + \ldots
\end{equation}
with $\hat{m}$ and $\langle \bar q q\rangle$ the renormalisation point
invariant current mass and condensate, respectively, and $d= 12/[33-2N_f]$;
$N_f$ is the number of quark flavours.  Comparing these two equations one
sees that, setting $d=1$ and neglecting $\ln[p^2]$ terms, the model defined
by Eqs.~(\ref{SSM}) and (\ref{SVM}) properly represents the ultraviolet
behaviour of the quark 2-point Schwinger function; incorporating asymptotic
freedom and dynamical chiral symmetry breaking, with
\begin{equation}
\label{cndst}
-\langle \bar q q \rangle = \frac{3}{4\pi^2} \frac{b_0}{b_1 b_3}~\lambda^3.
\end{equation}

Another feature of this model is that both $\sigma_S$ and $\sigma_V$ are
entire functions in the complex-$p^2$ plane but for an essential
singularity. As a consequence the quark 2-point Schwinger function does not
have a Lehmann representation and can be interpreted as describing a confined
particle.  This is because, when used in Eq.~(\ref{F}), for example, this
property ensures the absence of free--quark production thresholds, under the
reasonable assumptions that $\Gamma_\pi(p^2)$ is regular in the domain of
integration.  It follows from this that Eq.~(\ref{F}) is free of endpoint and
pinch singularities.

The expressions in Eqs.~(\ref{SSM}) and (\ref{SVM}) provide a six-parameter
model of the dressed quark 2-point Schwinger function in QCD: $C_m$, $\bar
m$, $b_0, \ldots, b_3$.  ($\Lambda = 10^{-4}$ is introduced simply to
decouple $b_2$ from the quark condensate.) These parameters can easily be
fitted to experimental observables, as we discuss below.

The Bethe--Salpeter amplitude, $\Gamma_\pi$ in Eq.~(\ref{F}), is the solution
of the homogeneous Bethe--Salpeter equation (BSE). Many studies of this BSE
suggest strongly that the amplitude is dominantly pseudoscalar (as already
indicated by the notation used in Eq.~(\ref{F})). Furthermore, in the chiral
limit the pseudoscalar BSE and quark DSE are identical\cite{DS79} and one has
a massless excitation in the pseudoscalar channel with
\begin{equation}
\label{gammapi}
\Gamma_\pi(p;P^2=0) = \frac{1}{f_\pi}\,B_{m=0}(p^2)~,
\end{equation}
where $B_{m=0}(p^2)$ is given in Eq.~(\ref{Sinv}) with $m=0$. This is the
realisation of Goldstone's theorem in the DSE framework; i.e., in the chiral
limit Eqs.~(\ref{SSM}) and (\ref{SVM}) completely determine $\Gamma_\pi$.

Herein we employ the approximation
\begin{equation}
\label{gammapim}
\Gamma_\pi(p;P^2=-m_\pi^2) \approx\frac{1}{f_\pi}\, B_{m=0}(p^2)~,
\end{equation}
which, for small current quark masses, is a good approximation both pointwise
and in terms of the values obtained for physical observables.\cite{FR95}

The quark-photon vertex, $\Gamma_\mu(p_1,p_2)$, satisfies a DSE that
describes both strong and electromagnetic dressing of the vertex.  Solving
this equation is a difficult problem that has only recently begun to be
addressed\cite{MF93}. However, much progress has been made in constraining
the form of $\Gamma_\mu(p_1,p_2)$ and developing a realistic
Ansatz\cite{DMR94,BP94}.  It is obvious that the bare vertex,
$\Gamma_\mu(p_1,p_2) = \gamma_\mu$, is inadequate when the fermion 2-point
Schwinger function has momentum dependent dressing because it violates the
Ward-Takahashi identity.  In Ref.~\cite{BC80} the following form was
proposed
\begin{eqnarray}
\Gamma_{\mu}^{\rm BC}(p,k) &  = &
\frac{\left[A(p^2) +A(k^2)\right]}{2}\;\gamma_{\mu} \nonumber \\
 & + &
\frac{(p+k)_{\mu}}{p^2 -k^2}\left\{ \left[ A(p^2)-A(k^2)\right]
                 \frac{\left[ \gamma\cdot p + \gamma\cdot k\right]}{2}
- i\left[ B(p^2) - B(k^2)\right]\right\} . \label{VBC}
\end{eqnarray}
This Ansatz is {\it completely determined} by the dressed quark 2-point
Schwinger function and has the features that it: 1) satisfies the
Ward-Takahashi identity, thereby ensuring current conservation at the
microscopic level; 2) is free of kinematic singularities - i.e., it has a
well defined limit as $p^2\to k^2$; 3) transforms correctly under $C$, $P$,
$T$ and Lorentz transformations; and 4) reduces to the bare vertex in the
manner prescribed by perturbation theory.  The fact that it is nevertheless
relatively simple makes it an ideal form to be employed in phenomenological
studies.  The studies of Ref.~\cite{R94} illustrate its phenomenological
efficacy.  In these studies, for example, it is shown that the chiral limit
for the anomalous decay $\pi^0\to\gamma\gamma$ is reproduced exactly using
\begin{equation}
\label{GTZ}
\Gamma_\mu(p,k) = \Gamma_\mu^{\rm BC}(p,k)~.
\end{equation}

{\bf 3. Chiral Limit: \mbox{\boldmath $\gamma \pi ^* \to \pi \pi $}.}
At the soft point in the chiral limit $(s=t=u=0)$ the transition form factor of
Eq.~(\ref{F}) is
\begin{eqnarray}
F^{3\pi}(0,0,0) = \frac{eN_c}{2\pi^2}
\int_0^\infty\,ds\,s\, \Gamma_\pi(s)^3\, \sigma_V\,
\Bigl\{  A\,\sigma_V\,\sigma_S
      &+& \case{3}{2} s A\, \sigma_V'\,\sigma_S
        - \case{3}{2} s A\, \sigma_V\,\sigma_S'
\nonumber \\
      & +&\case{1}{2} s A'\,\sigma_V\,\sigma_S
        - \case{1}{2} s B'\, \sigma_V^2 \Bigr\} ~.
\label{G0}
\end{eqnarray}
Defining
\mbox{$C(s) = B(s)^2/[s\,A(s)^2] = \sigma_S(s)^2/[s\,\sigma_V(s)^2]$}
one obtains a dramatic simplification and Eq.~(\ref{G0}) becomes
\begin{equation}
F^{3\pi}(0,0,0) = -\,\frac{eN_c}{2\pi^2}\int_0^\infty\,ds\,
        \frac{\Gamma_\pi(s)^3}{B(s)^3}\,\frac{C'(s)C(s)}{[1+C(s)]^4}~.
\end{equation}
Recalling Eq.~(\ref{gammapi}), which is the manifestation of Goldstone's
theorem in the DSE approach, it follows that
\begin{equation}
\label{gpigg0}
F^{3\pi}(0,0,0) = \frac{eN_c}{2\pi^2f_\pi^3}
\int_0^\infty\,dC\,\frac{C}{(1+C)^4} = \frac{eN_c}{12\pi^2f_\pi^3}~,
\end{equation}
since $C(s)$ is a monotonic function for $s\ge 0$ with $C(s=0)=\infty$ and
$C(s=\infty)=0$.  Hence, the chiral limit value \cite{ALTZ71} is reproduced
{\it independent} of the details of the quark 2-point Schwinger function,
$S(p)$.

In order to obtain the result in Eq.~(\ref{gpigg0}) it is essential that, in
addition to Eq.~(\ref{gammapi}), the photon-quark vertex satisfy the Ward
identity.  This is not surprising.  However, the fact that one must dress all
of the elements in the calculation consistently is often overlooked.  The
subtle cancellations that are required to obtain this result also make it
clear that it cannot be obtained in model calculations where an arbitrary
cutoff function (or ``form-factor'') is introduced into each integral.  The
fact that the pion Bethe-Salpeter amplitude is proportional to the scalar
part of the quark self energy in the chiral limit, Eq.~(\ref{gammapi}), is
crucial. These features are also to be seen in the calculation\cite{PRC87} of
the Wess-Zumino five pseudoscalar term and the $\pi^0\gamma\gamma$
vertex\cite{R94}, in which again the values expected from considerations of
anomalous current conservation are obtained independent of the details of the
quark 2-point Schwinger function, $S(p)$.

{\bf 4. Numerical Results.}
The chiral limit value of the  $\gamma\pi\pi\pi$ amplitude, $F^{3\pi}(0,0,0)$,
provides a useful normalisation.  We therefore define the function
\begin{equation}
\tilde F^{3\pi}(s,t,u) = F^{3\pi}(p_1,p_2,p_3)/F^{3\pi}(0,0,0) =
\case{1}{e}\,4\pi^2 f_\pi^3\, F^{3\pi}(p_1,p_2,p_3)~.
\label{Ft}
\end{equation}
We employ the following definition of the Mandelstam variables:
\begin{equation}
s=-(p_1+p_2)^2\equiv m_\pi^2 \bar s, \quad t=-p_3^2\equiv m_\pi^2\bar t, \quad
u=-(p_1+p_3)^2~\equiv m_\pi^2\bar u,
\label{stu}
\end{equation}
which ensures that even though we work in Euclidean metric these variables
have their conventional interpretation.  We note that the quantity
$t-m_\pi^2$ provides a measure of the amount by which the (third) pion is
off-shell.

In the experiment proposed at CEBAF the photon energy in the proton rest
frame is between 1 and 2 GeV.  This suggests the following range of
Mandelstam variables: $4 \le \bar s \le 16$, $-9 \le \bar t \le -1$, $-16 \le
\bar u \le 5$.  For the purpose of the numerical calculation we fix $\bar t=
- 1$; i.e., we choose pion 3 to be as close as (experimentally) possible to
its mass shell, $\bar t=1$. Within the range of $s$ considered, the
requirement of a fixed photon energy of 1 GeV in the proton rest frame
entails \begin{equation} \bar u=(-1.5-4.28\,x+0.012\,x^2) \quad {\rm with}
\quad x=(\bar s/4 -1).  \label{ufit} \end{equation}

We have demonstrated above that in the chiral limit \mbox{$F^{3\pi}(0,0,0)= e
N_C/(12\pi^2 f_\pi^3)$} independent of the model parameters.  The evolution
of $F^{3\pi}(s,t,u)$ with $s$ does depend on the model parameters, even in
the chiral limit.  As we have described above, Eqs.~(\ref{SSM}), (\ref{SVM}),
(\ref{gammapi}) and (\ref{GTZ}) provide a six-parameter model [$C_m$,
$\overline{m}$, $b_0\ldots b_3$]of the nonperturbative, dressed-quark
substructure of the pion based on DSE studies.  These parameters are fixed by
requiring that the model reproduce, as well as possible, the following pion
observables \mbox{$f_\pi/\langle \overline{q}q\rangle_{1\,{\rm GeV}^2}^{1/3}
= 0.42\pm 0.02$}, \mbox{$f_\pi\,r_\pi = 0.31\pm 0.01$},
\mbox{$m_\pi^2/\langle \overline{q}q\rangle_{1\,{\rm GeV}^2}^{2/3} = 0.40 \pm
0.03$}; the dimensionless $\pi$-$\pi$ scattering lengths (discussed in
Refs.~\cite{RCSI94,Poc95,MES94} with current experimental values presented in
Table~\ref{table}) and the pion electromagnetic form factor on
spacelike-$q^2\in [0,4]$~GeV$^2$\cite{R94}.

The values of observables are given by simple integral expressions involving
the quark 2-point Schwinger function and pion Bethe-Salpeter amplitude; for
example,\cite{FR95}
\begin{eqnarray}
\lefteqn{f_\pi^2=}\\
& &
 \frac{N_c}{8\pi^2}\int\,dp^2\,p^2\,B_{0}(p^2)^2\,
\left(  \sigma_{V}^2 - 2 \left[\sigma_S\sigma_S' + s
\sigma_{V}\sigma_{V}'\right]
  - s \left[\sigma_S\sigma_S''- \left(\sigma_S'\right)^2\right]
- s^2 \left[\sigma_V\sigma_V''- \left(\sigma_V'\right)^2\right]\right)~;
\nonumber
\end{eqnarray}
\begin{equation}
\label{expmass}
m_\pi^2\,f_\pi^2 =
\frac{N_c}{2\pi^2}\,\int\,dp^2\,p^2\,
        \frac{B_{0}(p^2)}{B(p^2)}
    \left(B(p^2)\,\sigma_S^{0}(s) - B_{0}(p^2)\,\sigma_S(p^2)\right)~,
\end{equation}
which follows from the pion Bethe-Salpeter equation and is consistent with
Dashen's relation\cite{Dashen69}; and
\mbox{$\langle \overline{q}q\rangle_{\mu^2}
= \ln\left[\mu^2/\Lambda_{\rm QCD}^2\right]\langle \overline{q}q\rangle$}
with $\langle \bar q q\rangle$ given in Eq.~(\ref{cndst}) and $\Lambda_{\rm
QCD}=0.2$~GeV.  In these equations the subscript or superscript ``$0$''
indicates that the labelled function has been evaluated with zero quark
current mass.

Following this procedure one obtains
\begin{eqnarray}
& &
\begin{array}{lll}
C_0=0.121, & \overline{m} = 0.00897~, & C_{\bar m} = 0,
\end{array} \nonumber \\
& &
\begin{array}{llll}
b_0 = 0.131~, & b_1 = 2.90~, & b_2 = 0.603 ~,& b_3 = 0.185~.
\end{array}
\label{ParamV}
\end{eqnarray}
The mass scale is set by requiring that $f_\pi$ take its experimental value
of $92.4$~MeV, which yields $D=0.160$~GeV$^2$.  The calculated values of
observables are presented, for completeness, in Table~\ref{table}.  The form
factor is discussed in Refs.~\cite{R94,BRT95}.

The amplitude in Eq.~(\ref{F}) is given by a four-dimensional integral that
we evaluate by straightforward Gaussian quadrature.  The calculation is
simplified by working in the rest-frame of one of the outgoing pions.  All
results reported here are tested against variation of numerical parameters,
such as grid densities, etc.

The quark 2-point Schwinger function specified by Eqs.~(\ref{Sp}),
(\ref{SSM}) and (\ref{SVM}) is an entire function.  The quantity
$x[\bar\sigma_V^0(x)]^2 + [\bar\sigma_S^0(x)]^2$ has a pair of complex
conjugate zeros at \mbox{$x=-0.102\pm i\,0.715$}, which corresponds to
$|p^2=x/(2 D)| \simeq 12\,m_\pi^2$.  This affects our calculation through the
approximation of Eq.~(\ref{gammapim}) because $B_0(p^2)$ has poles at these
points. These poles are integrable singularities and do not lead to an
imaginary part in the amplitude.  However, in our calculation of
$F^{3\pi}(s,t,u)$ we treat them as a spurious artifact of the approximation
of Eq.~(\ref{gammapim}) and consider our results to be unreliable when these
poles enter the integration region.  This occurs only for $\bar s > 12$.  The
approximation of Eq.~(\ref{gammapim}) has hitherto only been tested for real,
spacelike-$p^2$ and the present application involves a considerable
extrapolation into the complex-$p^2$ plane as $\bar s$ is increased.  (In the
chiral limit, $m_\pi^2=0$, the integration does not explore the complex-$p^2$
plane and these poles do not enter the integration region.)  The importance
of this observation is that processes such as $\gamma\pi^\ast\to\pi\pi$
provide a means of exploring the structure of bound state amplitudes in the
complex-$p^2$ plane.

\begin{figure}
  \centering{\
     \epsfig{figure=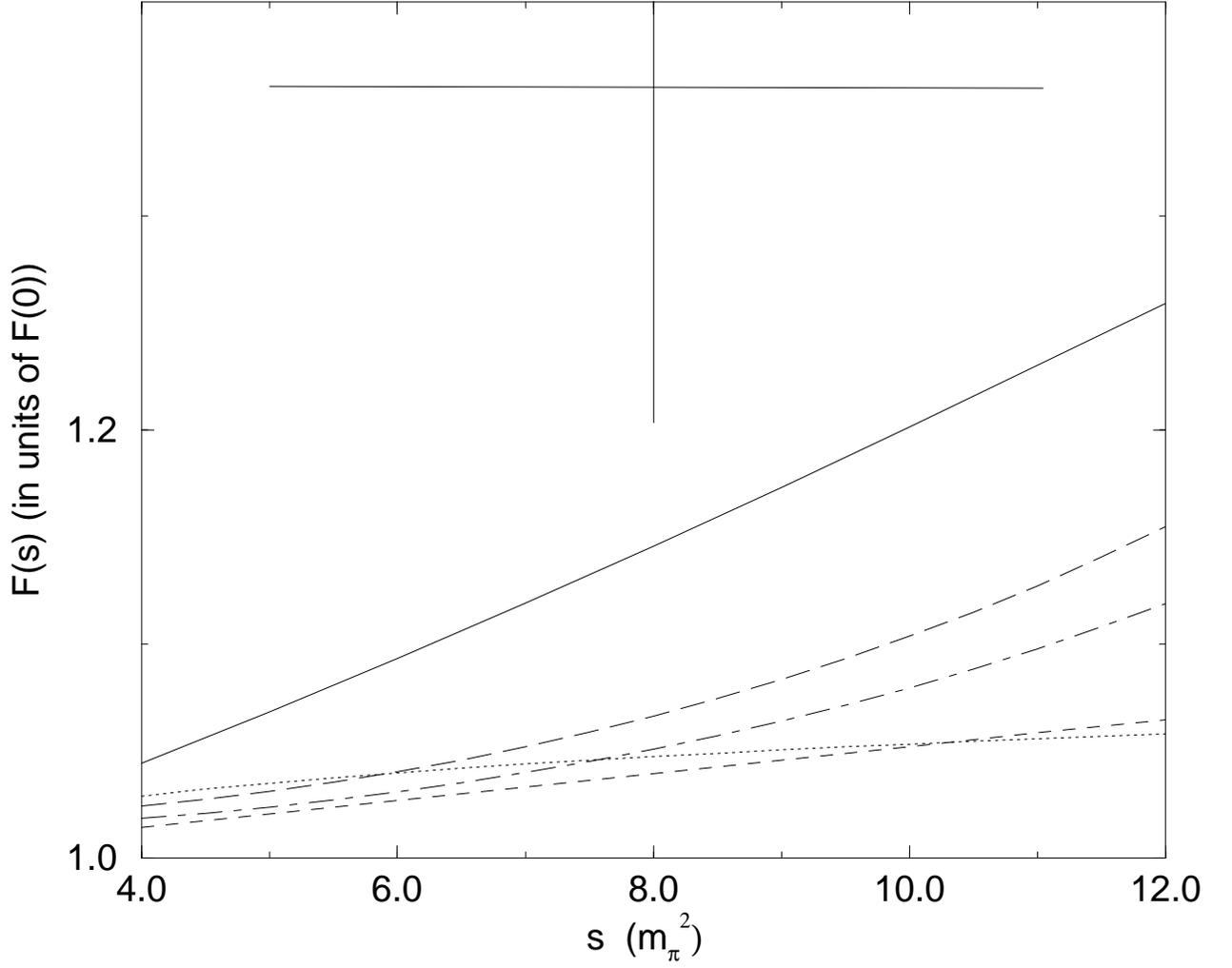,height=16cm,rheight=16cm}  }
\caption{The normalized $\gamma \pi ^* \pi \pi $ amplitude $\tilde
F^{3\pi}(s,t,u)$, Eq.~(\protect{\ref{Ft}}), as function of the Mandelstam
variable $\bar s$ for different values of $\bar u$ as given by
Eq.~(\protect{\ref{ufit}}) at $\bar t=-1$.  Our result: solid line; our
result with $m_\pi \to m_\pi/2$: short-dash line; vector meson dominance:
long-dash and dash-dot lines, see Eqs.~(\protect\ref{VMD2}) and
(\protect\ref{VMD1}); chiral expansion plus vector meson saturation Ansatz:
dotted line, see Eq.~(\protect\ref{ChPT}).  The data point is taken from
Ref.~\protect\cite{Antipov}.
\label{figFs} }
\end{figure}

Our numerical results are shown in Fig.~\ref{figFs}. The solid line is our
calculated result for the $\gamma \pi ^* \pi \pi $ amplitude $\tilde
F^{3\pi}(s,t,u)$, Eq.~({\ref{Ft}}), as function of the Mandelstam variable
$\bar s$ for different values of $\bar u$ as given by Eq.~(\ref{ufit}) at
$\bar t=-1$.  An excellent fit to this curve is given by
\begin{equation}
\tilde F^{3\pi} (s,t,u)
=1.044 + 0.096\, x + 0.006\, x^2 \quad {\rm with} \quad  x=(\bar s/4-1).
\label{Ffit}
\end{equation}
Fixing the Mandelstam variables such that the energy of the incident photon
is 2 GeV in the proton rest-frame changes our curve by an amount that is not
visible on the scale of this plot.  Importantly, the result is {\it
insensitive} to the details of the parametrisation of the quark 2-point
Schwinger function.  A quantitatively similar curve is obtained using earlier
sets of the parameters in Eq.~(\ref{ParamV}); i.e., our result is not
sensitive to the details of the model.  The form factor does depend on the
pion mass.  The short-dashed line is the result we obtain with $m_\pi =
m_\pi^{\rm exp}/2$.

A comparison of our result with that obtained in other models reveals that
all results are broadly consistent. Our prediction for this form factor is,
however, uniformly larger and displays a more rapid increase with $\bar s$.
For example, using vector meson dominance one obtains, for real
photons\cite{VMD2},
\begin{equation}
\tilde F^{3\pi} (s,t,u) = 1+ C_\rho e^{i\delta }
\left( \frac s {m_\rho^2 -s} + \frac {t'}{m_\rho^2 - t' } + \frac
u {m_\rho^2 - u} \right)
\label{VMD2}
\end{equation}
with $t'=s-t+u-2m_\pi^2$, $C_\rho = 2g_{\rho \pi\pi}g_{\rho \pi\gamma}/
[m_\rho^3 F^{3\pi} (0,0,0)] \approx 0.434$ and $\delta$ an unknown phase.
This expression, with $\delta=0$, appears as the long-dashed line in
Fig.~\ref{figFs}.  This curve is smaller in magnitude than our result for all
$\bar s$ and rises more slowly with $\bar s$.  Another vector meson dominance
model estimate of this process\cite{VMD1} appears as the dash-dot line in
Fig.~\ref{figFs}.  This result is obtained from
\begin{equation}
\tilde F^{3\pi} (s,t,u) = \frac {m_\rho^2} 3
\left( \frac 1 {m_\rho^2 -s} + \frac 1 {m_\rho^2 - t' } + \frac
1 {m_\rho^2 - u} \right).
\label{VMD1}
\end{equation}
and is even smaller in magnitude and has a weaker $\bar s $ dependence than the
other vector meson dominance estimate.

The weakest $\bar s$ dependence is obtained using a model based on chiral
expansion techniques and employing a vector meson saturation
Ansatz\cite{ChPT}.  This result appears as the dotted line in
Fig.~\ref{figFs} and is obtained from the expression
\begin{equation}
\tilde F^{3\pi} (s,t,u) = \left|1 + C_{\rm pion-loops} +
\frac {s+t'+u}{2m_\rho^2}\right|
\label{ChPT}
\end{equation}
where the non-divergent part of the coefficient $C_{\rm pion-loops}$ is
\begin{equation}
C_{\rm pion-loops} = \frac 1{96\pi^2f_\pi^2}
\left[ \left(\log\left[\frac {m_\rho^2}{m_\pi^2}\right]+ \frac
53\right)(s+t'+u)
+ 4m_\pi^2 \left\{f(s)+f(t')+f(u)\right\} \right]
\label{Cloop}
\end{equation}
with $f(\zeta<0)$ defined by
\begin{equation}
f(\zeta)=(1-\zeta/4m_\pi^2)\sqrt{1-4m_\pi^2/\zeta}
\log \left( \frac {\sqrt{1-4m_\pi^2/\zeta}+1}{\sqrt{1-4m_\pi^2/\zeta}-1}
\right) -2~.
\end{equation}
Analytic continuation is used to define $f(\zeta > 0)$.  Note that $f(\zeta)$
has an imaginary part for $\zeta > 4m_\pi^2$; i.e., for $s$ in the domain
explored by the existing and proposed experiments.  The imaginary part is due
to the pion loop in the $s$--channel.

We can compare our result with the one existing data point\cite{Antipov}. Its
statistical and systematic error as well as the uncertainty in $s$ is also
displayed in Fig.~\ref{figFs}. The fact that this data point is well above
the chiral limit prediction has caused some concern \cite{Exp1}. However,
given the experimental errors and the prediction of our model this data point
does not appear untenable.

{\bf 5. Summary and Conclusions.} Herein we have reported a calculation of
the form factor for the anomalous process $\gamma \pi ^* \to \pi \pi $,
$F^{3\pi}(s,t,u)$, in a phenomenological approach based on the QCD
Dyson--Schwinger equations (DSEs). In our approach the chiral limit value of
$F^{3\pi}(0,0,0)$ is reproduced {\it independent} of the details of the quark
2-point Schwinger function.  Using for the $u$-$d$-quark 2-point Schwinger
function a parametrisation fitted to low--energy pion data, $F^{3\pi}(s,t,u)$
was calculated for a range of $(s,t,u)$ specified in such a way as to cover
the kinematic region to be explored in a proposed experiment\cite{Exp1}.  The
small photon virtuality in another proposed experiment\cite{Exp2} should also
make it possible for our prediction to be compared, without adjustment, with
the results of that experiment.

We compared our result with that obtained in other models when applied in the
same $(s,t,u)$ range as we have considered, which is the range appropriate
for the experiment proposed at CEBAF\cite{Exp1}.  Our model yields a form
factor, $F^{3\pi}(s,t,u)$, that is uniformly larger and has a more rapid
increase with $s$ than any of the other models considered.  The result
obtained in a model based on chiral expansion techniques, augmented by vector
meson saturation assumptions, leads to the weakest $s$ dependence.  Although
the models are broadly consistent, the variation between the results is such
that the proposed experiments should be able to distinguish between them.

Along with the study in Ref.~\cite{FMRP95}, this phenomenological application
of the QCD DSEs is one of the first to explore the model quark 2-point
Schwinger function and pion Bethe-Salpeter amplitude in the timelike region,
which is not accessible in perturbation theory.  This region is important in
the study of, for example, vector meson Bethe-Salpeter and baryon Fadde'ev
amplitudes.  Experimental data on $F^{3\pi}(s,t,u)$ can therefore be used to
place additional constraints on the analytic structure of the QCD Schwinger
functions.

\acknowledgements
This work was supported by the US Department of Energy, Nuclear Physics
Division, under contract number W-31-109-ENG-38 and by the Deutsche
Forschungsgemeinschaft (DFG) under contract number AL~297/2-1.  The
calculations described herein were carried out using a grant of computer time
and the resources of the National Energy Research Supercomputer Center.
R.A.\ thanks the Physics Division of ANL for their warm hospitality during
two visits in which most of the work described herein was performed and Prof.\
H.\ Reinhardt for his support.

%______________________________ References ______________________________

%____________________________________________________________________________
\begin{table}
\begin{tabular}{|c|l|l|}
   & Calculated  & Experiment  \\ \hline
  $f_{\pi} \; $    &  ~0.0924 GeV &   ~0.0924 $\pm$ 0.001     \\ \hline
  $m_{\pi} \; $    & ~0.1385  & ~0.1385  \\ \hline
  $m^{\rm ave}_{1\,{\rm GeV}^2}$ & ~0.0051 & ~0.0075
                \\ \hline
 $-\langle \bar q q \rangle^{\frac{1}{3}}_{1\,{\rm GeV}^2}$ & ~0.221 &
        ~0.220\\ \hline
 $r_\pi \;$ & ~0.55 fm & ~0.663 $\pm$ 0.006  \\  \hline
 $g_{\pi^0\gamma\gamma}\;$ & ~0.505 (dimensionless) & ~0.504 $\pm$
0.019\\\hline
 $F^{3\pi}(4m_\pi^2)\;$ & ~$1.04$ & ~$1$ (Anomaly)\\\hline
 $a_0^0 \;  $ & ~0.17  & ~0.21 $\pm$ 0.01 \\ \hline
 $a_0^2 \;  $ & -0.048 & -0.040 $\pm$ 0.003 \\ \hline
 $a_1^1 \;  $ & ~0.030 & ~0.038 $\pm$ 0.003\\ \hline
 $a_2^0 \; $  & ~0.0015 & ~0.0017 $\pm$ 0.0003\\ \hline
 $a_2^2 \;  $ & -0.00021 & \\ \hline
\end{tabular}
\caption{Pion observables calculated using the parameter values in
Eq.~(\protect\ref{ParamV}).  The ``experimental'' values listed for $m^{\rm
ave}$ and $\langle \bar q q \rangle$ are an indication of other contemporary
theoretical estimates.  Experimental values not discussed in the text are
taken from Ref.~\protect\cite{PDG94}. The difference between the calculated
and experimental values of $r_\pi$ is a measure of the importance of
final-state $\pi$-$\pi$ interactions and photon-$\rho$-meson
mixing\protect\cite{ABR95}; that between the calculated and experimental
values of the pion scattering lengths is a measure of the importance of
$\pi$-$\pi$ final-state interactions in this case\protect\cite{RCSI94}.
\label{table}}
\end{table}
%____________________________________________________________________________

\end{document}